# Dark Matter with (very) heavy SUSY scalars at ILC

M. BERGGREN[a], F. RICHARD[b], Z. ZHANG[b,1]

[a]Laboratoire de Physique Nucléaire et de Hautes Energies,
IN2P3-CNRS et Université de VI-VII, 4 place Jussieu, Tour 33, 75252 Paris Cedex 05 France

[b]Laboratoire de l'Accélérateur Linéaire,
IN2P3-CNRS et Université de Paris-Sud XI, Bât. 200, BP 34, 91898 Orsay Cedex France

**Abstract**

In this paper, six SUSY scenarios with heavy sfermions, mainly based on theoretical arguments and on experimental indications for new physics, are defined. These scenarios, consistent with the amount of dark matter (DM) measured by WMAP, are then analysed in detail providing pertinent examples of the potential of ILC. It is shown that in most cases ILC, with its high precision based on the chargino analysis and in spite of an incomplete coverage of the gaugino and slepton mass spectrum, can predict the amount of DM in our universe with an accuracy which matches the WMAP results.

---

[1] E-mail: mikael.berggren@cern.ch, richard@lal.in2p3.fr, zhangzq@lal.in2p3.fr

# I Introduction

The present common belief, based on the very precise WMAP [1] results, is that 85% of the mass of the universe is made out of Dark Matter, DM. SUSY provides the best candidate(s) so far to explain DM but it remains to be proven that the lightest neutralino is the main source of this DM.

This quest is already undertaken by non-accelerator experiments and could be the main goal of LHC and ILC in the next 10-15 years.

Non-accelerator experiments could detect (have already detected?) a weakly interacting massive particle (WIMP) signal, but they cannot provide an accurate comparison to WMAP since their results depend on the product of the local density and on the WIMP cross section. Even assuming that one can predict the former and therefore estimate the later, this observable alone plus a crude estimate of the WIMP mass are insufficient to match the precise measurement of WMAP.

LHC should provide the first proof of existence of SUSY and give precious indications on the neutralino properties and mass which will constitute a solid basis to assess the nature of DM.

Is this sufficient? The answer is definitely no since, even in a favourable scenario, LHC cannot provide an adequate accuracy to match the WMAP, and even more, the PLANCK accuracies (figure 1).

Why is it so important to reach such accuracies? Firstly, if there is coincidence between the two results, one will establish that the neutralino is indeed the main, presumably the only, source of DM in the universe. This result is certainly non-trivial since there may be additional sources of DM, e.g.:

- within the Standard Model (SM) axions
- within the neutrino sector neutrino condensates (with spin-statistics violation as in [2])
- within SUSY axinos, Q-balls (squark condensates), gravitinos which can decay into neutralinos
- beyond SUSY SM Kaluza Klein excitations, inflationary objects ('wimpzillas', 'cryptons').

One could alternatively find that the results of colliders over-close DM which could mean that the neutralino is long-lived but unstable and decays into a gravitino which constitutes the true DM.

While the precise estimate of DM relies on the lightest part of the SUSY spectrum, accessible at ILC, some control on the heaviest part of the spectrum is needed given the high accuracy required. We will see how precise polarisation measurements at ILC can provide the essential missing parts of the SUSY parameters and reduce the model dependence of the collider results.

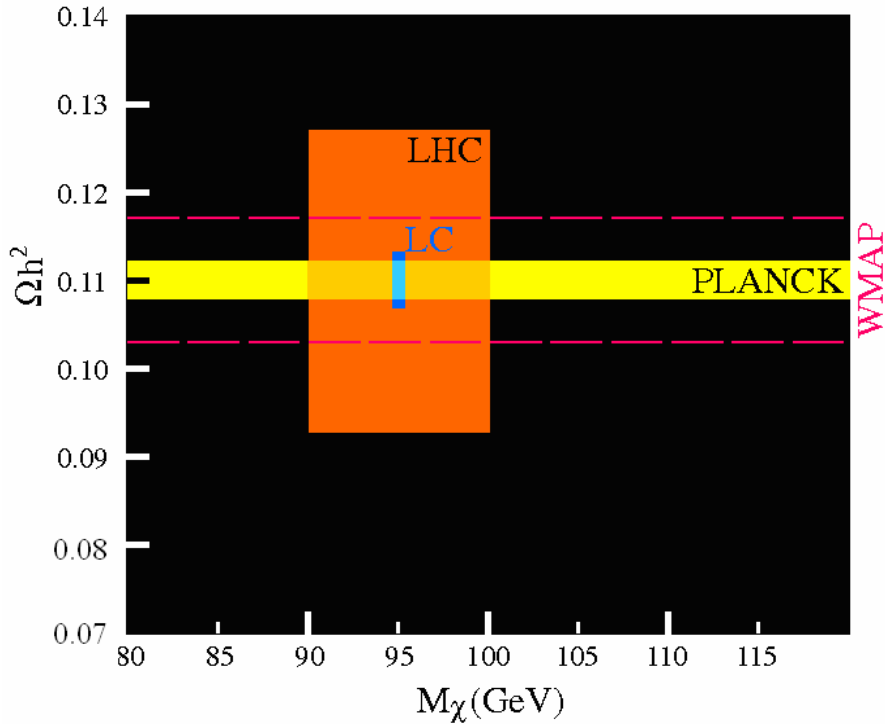

**Figure 1:** *This figure illustrates the expected accuracy of ILC [3] and LHC for predicting the amount of dark matter in the universe by analysing the SUSY data observed at these colliders. The chosen point is relatively friendly for what concerns LHC. In comparison are shown the WMAP and the expected PLANCK SURVEYOR accuracies.*

A word of caution is necessary when discussing the ultimate reachable accuracies. What ILC/LHC can only provide is a spectrum of SUSY particles and an estimate of the cross sections relevant for DM annihilation. This is just part of the problem since one also needs a precise modelling of the cooling process in the primordial universe. This issue is clearly outside the scope of our field but should not be forgotten when estimating the final accuracy.

In a previous paper [3], the so-called 'co-annihilation' scenario has been investigated in detail with the result that, in most cases, ILC was able to match the Planck accuracy. This is shown pictorially in figure 1 which takes as a reference an 'easy' SUSY solution for which LHC can also produce a valuable measurement.

The purpose of the present paper is to examine different SUSY scenarios, in which all the sfermions, scalar partners of SM fermions, are heavier than 1 TeV (and in some instances much heavier) and therefore the only relevant particles are the lightest gauginos which could be detected at ILC. ILC should also allow testing the absence of light sfermions through precise polarisation measurements as will be shown in the various scenarios under consideration.

The motivations for such scenarios will be recalled in the next section. In particular, some issues on the flavour problem will be recalled, like the stringent limits on FCNC and CP violation which come from the electric dipole moment (EDM) of electrons and neutrons. These limits set very severe constraints on phases of the SUSY parameters which would have to be chosen null or unnaturally small. With very heavy sfermions one can allow for natural phases.

In section I six working points, chosen according to these motivations, will be described while in section III experimental problems related to these points are discussed. Section IV will give the results of this analysis which will be used in section V to deduce ILC accuracies on DM.

## II The heavy sfermion scenarios

Various arguments: FCNC, EDM and the proton lifetime ($\tau_p$), lead to the idea that SUSY phenomenology favours heavy scalars. How heavy? Fine-tuning (FT) is usually advocated as a limiting factor but various authors point out that SUSY is already finely tuned 'at the 1%' level.

The most stringent constraints on phases are coming from EDM limits on electrons and neutrons which are continuously progressing [4] as can be seen in figure 2.

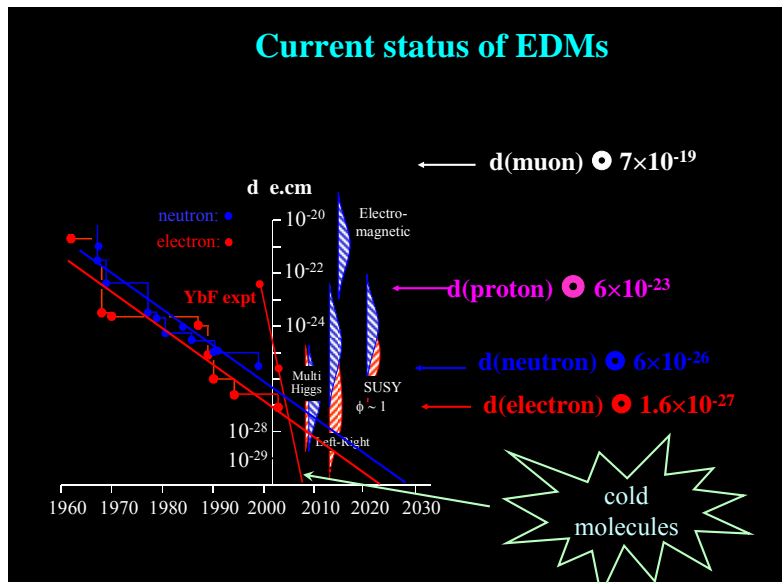

**Figure 2:** *This figure, taken from [4], illustrates the rapid progress on neutron and electron EDM sensitivities achieved in recent experiments. It also shows that using molecules, this progress is becoming even faster allowing to cover most theoretical predictions.*

This picture illustrates qualitatively the trend. Progress on these limits has recently accelerated with the advent of a new generation of experiments using cold molecules. In the SM EDM expectations are at the $10^{-35}$ e.cm level hence negligible with respect to SUSY expectations if SUSY phases are large as for instance in the CKM matrix.

There are two possibilities to avoid these constraints:

- SUSY phases are 0 from an underlying principle, possibly CP conservation within SUSY, yet to be justified

- SUSY particle masses (non necessarily all of them) are very large, well beyond the LHC/ILC reach.

While the first hypothesis seems to allow the survival of mSUGRA with low FT, it does not take care of the FCNC issue [5]. The second one maintains the possibility of 'naturally' large phases and FCNC assuming that the sfermions are heavier than 10 TeV.

A more radical idea, discussed in the following, is the Split SUSY scheme [6]. Even then, one finds that two-loop corrections involving only the light SUSY fermions predict a measurable EDM contribution as shown in figure 3.

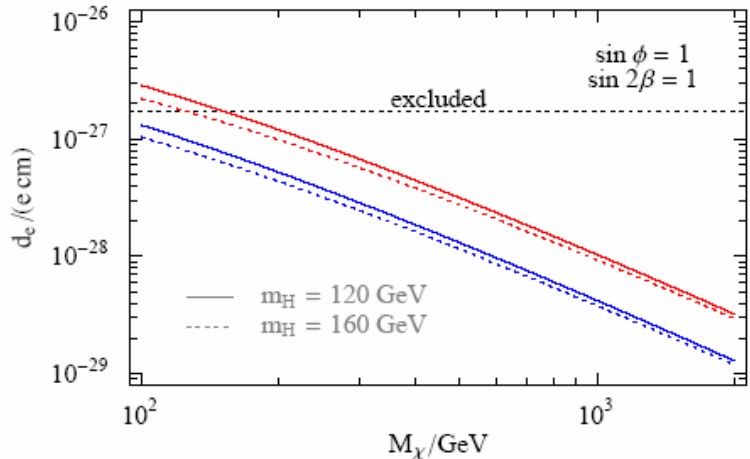

**Figure 3:** *Curves from [6] showing that in the SpS scheme, the EDM prediction with large CPV phases will be tested soon.*

**mSUGRA**, while being the simplest reference to discuss collider phenomenology and passing most the experimental constraints including DM, does not provide a compelling mechanism to solve the flavour problems unless fermion masses are taken heavy (the **Focus** solution described below), hence the other solutions proposed: gauge-mediated, gaugino-mediated and anomaly mediated SUSY breaking schemes which claim to do a better job. The latter, referred as **AMSB**, claims complete UV decoupling, which means that SUSY-breaking is not influenced by the mechanisms which generate flavour symmetry breaking (at high energy scale), and in particular there is no extra CP violation as compared to the SM. In AMSB one can also accommodate a $g-2$ contribution since there could be light sleptons. Also AMSB does not necessarily impose the mSUGRA hierarchy between the gaugino masses with the condition $M_1 \sim 0.5 M_2$, but allows the possibility that the LSP is a wino which allows a much heavier LSP.

A different attitude, recently emphasized [6], would be to ignore the FT criteria given that our world appears extremely fine-tuned. One could therefore completely give up SUSY at the TeV scale and forget it at LHC-ILC energies or, less radically, only keep the 'good part' of SUSY, the fermions at ~1 TeV, which give us GUT and DM and send the troublesome scalars (except the light Higgs boson naturally light) to very high scales. This approach, which is based on string theory arguments (a continuum of cosmological solutions, our universe being a very particular one with galaxies and carbon nuclei) is known as **Split SUSY** and will be referred as **SpS**. This scenario will be discussed in more detail but it is worth mentioning that:

- it gets rid of the severe constraints set by the EDM limits on the phases of SUSY parameters ($\mu$ and $M_1$)

- from DM constraints, $\mu$ and $M_1$ are tightly correlated as shown in figure 4

- if one assumes very massive scalars, say $M_S$>1000 TeV, one has to take into account some running effects which modify the non-diagonal terms in the gaugino mass matrices which therefore not only depend on tan$\beta$ but also on $M_S$. In fact it turns out that the precise measurements performed at ILC may allow to have a handle on $M_S$.

From the various schemes discussed above, one can distinguish between 3 possible regimes.

1/ If $M_1$ is much larger than $\mu$, the LSP is a pure Higgsino and therefore one needs $\mu$~1 TeV to saturate the WMAP constraint. This type of solution corresponds to a TeV chargino almost degenerate with the LSP and the NLSP. It would require a 2 TeV LC to produce the lightest chargino. In this paper we will show that the detection of the chargino is possible at a LC. Squarks and gluinos being very heavy, LHC would not observe SUSY. Charginos would have to be produced through the Drell-Yan process but could not be observed since they are almost degenerate with the LSP and the cross sections are too small.

2/ If $\mu$~$M_1$, then the LSP can be as light as 50 GeV (LEP2 constraint). The correlation between these two parameters is displayed in figure 4 and is very similar in the SpS and Focus solutions discussed below.

3/ If $\mu$ is much larger than $M_1$ and $M_2$ one can still reproduce WMAP by relaxing the GUT constraint between $M_1$ and $M_2$, taking e.g. $M_2$=2 TeV and $M_1$ much larger than $M_2$ (wino regime, which does not occur within SUGRA but is possible within AMSB). This regime requires a LC even beyond 4 TeV and appears also hopeless at LHC.

Scenarios 1 and 3 may seem quite desperate for future colliders, but it must be said that they would correspond to seriously fine-tuned SUSY solutions. It should also be recalled that one needs not saturate the WMAP DM measurement by the sole neutralino component since, as mentioned, there are several other sources of DM. For this reason, we have retained a type 1 solution accessible at ILC and selected mostly type 2 solutions.

We may now proceed to the description of the solutions retained in this paper. They have been selected with the following criteria:

- they correspond to heavy sfermions (all)
- they rely on theoretical schemes trying to solve the flavour problem (1, 2, 3, 6)
- they are inspired by some experimental indications (4, 5)
- they have interesting detection features (6).

The following SUSY codes have been used:

- SuSpect [7] to generate a consistent set of SUSY parameters fulfilling the EWSB constraint
- Susygen [8] to generate the masses of SUSY gauginos and compute cross sections
- Micromegas [9] to compute the amount of DM.

**The Focus solution**

The Focus mechanism can be worked out within mSUGRA, for values of $m_0$ as large as 10 TeV satisfying the EWSB constraint and with acceptable FT. Since $M_1$~0.5$M_2$, the LSP is a bino with insufficient annihilation rate to generate the right amount of DM. It is possible however to solve this problem by having $\mu$ smaller or at least of the same order as $M_1$, the so-called 'Higgsino-gaugino' mixing regime.

Using SuSpect, from EWSB one finds $|\mu|$=430 GeV with $\tan\beta$=10, $A_t$=0, $m_{1/2}$=900 GeV, $m_0$=12.5 TeV and sign($\mu$)>0 consistent both with WMAP and with EWSB. This high value of $m_0$ removes any restriction on the phase of $\mu$ due to the electron EDM limits. Note also that this solution corresponds to a 178 GeV top mass, compatible with the 2004 value. Had we chosen 174 GeV, today's value, the preferred $m_0$ value would be significantly lower.

**The SpS solutions**

These solutions are similar to the Focus solution, but with much heavier sfermions. It also requires a tight correlation between $\mu$ and $M_1$ to generate the proper amount of DM.

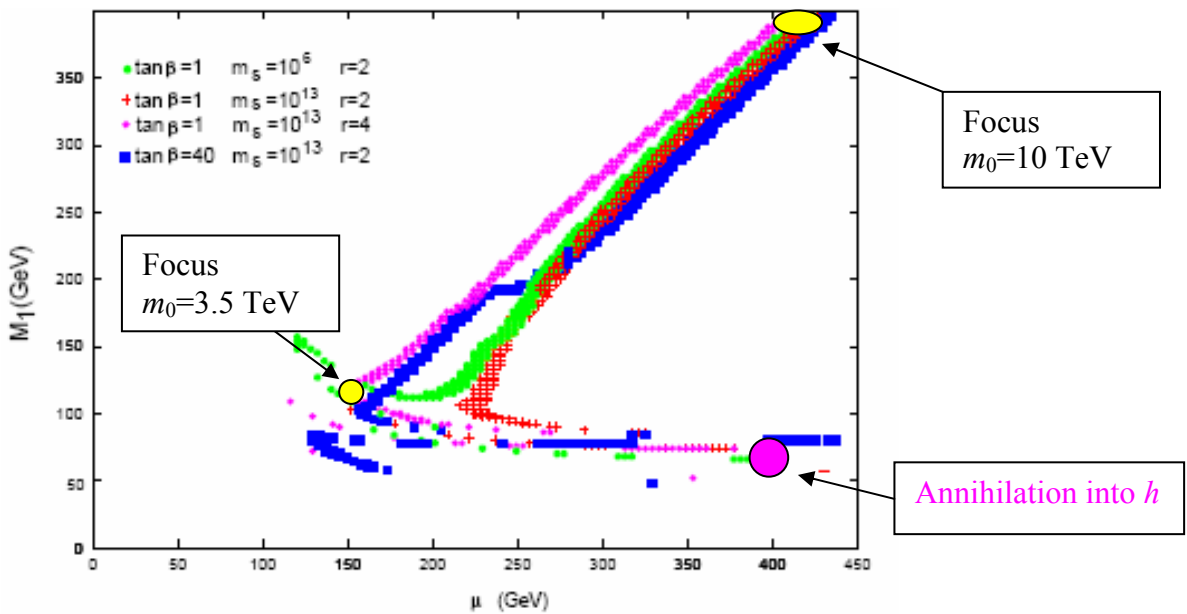

**Figure 4:** *Plot from [10] showing the tight correlation between the gaugino mass parameter $M_1$ and the Higgsino mixing mass $\mu$ for the SpS and the Focus solutions discussed in the text. They are similar with gaugino-Higgsino mixing needed to generate the proper amount of DM. They have scalar masses which differ by several orders of magnitude. The horizontal line corresponds to the possibility to annihilate light binos through the lightest Higgs scalar which, in SpS, can be as heavy as 170-180 GeV.*

In addition, if $m_\chi \sim m_h/2$, the Higgsino component can be much smaller ($\mu \gg M_1$) since annihilation through the Higgs pole becomes very large. Note that the rate of annihilation is very dependent on the speed distribution of the WIMPs since the annihilation process is *p*-wave (contrary to the annihilation process in the EGRET solution (see below) which proceeds through an *s*-wave into the CP odd *A* boson). Recently [11] this scenario has been emphasized also within mSUGRA with of course much less fine-tuning.

In summary, two choices have been retained as shown in table I:
- an SpS solution with Higgsino-wino mixing tuned to the WMAP solution

- a bino solution where the LSP annihilates through *h*.

**Data inspired solutions: EGRET and LEP**

These solutions have been suggested by the interpretation in terms of SUSY of two reported excesses:

- an excess of photons from the satellite experiment EGRET
- an excess of events at LEP2 interpreted as due to the Higgs SUSY sector.

**The EGRET solution**

In present direct or indirect searches various indications of excesses above standard expectations have been reported. None of them stands for itself and, as suggested in [12], it will require a combination of indications with a common interpretation to take seriously such results. Only with the advent of collider data will one be able to reach full consistency and filter out wrong interpretations. In spite of these uncertainties we have chosen to take seriously one of these indications and its detailed interpretation in terms of SUSY.

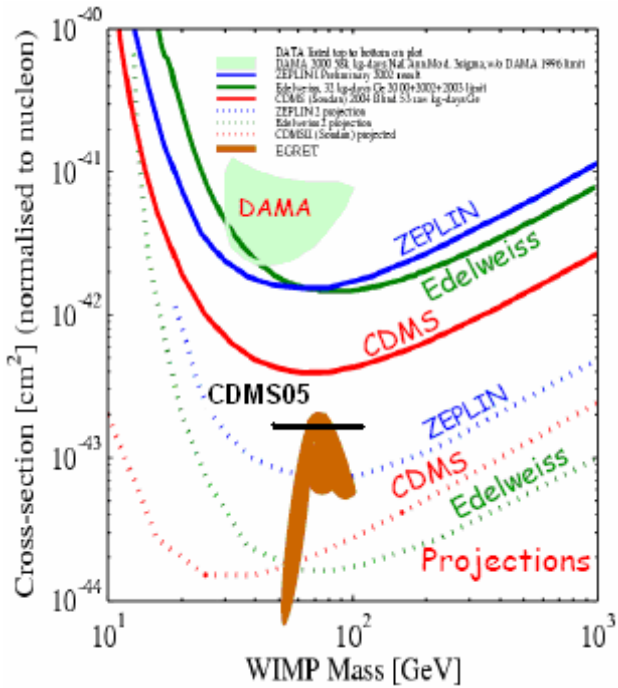

**Figure 5:** *The brown spot in this figure from [15] indicates the expected WIMP mass and cross section for direct detection in the EGRET scenario. The dark line from [16] shows that CDMS present sensitivity comes quite close to the expected signal.*

From the analysis of photon data taken with EGRET [13], an excess has been reported with respect to standard expectations. The significance of this excess can be taken seriously, in spite of the uncertainties on the modelling of [14] of the background, for two reasons. Firstly one observes a difference in shape and not just in normalisation which suggests that a different mechanism is at work. Secondly the excess shows that angular dependences can be interpreted in terms of the expected DM content of our galaxy [15].

The SUSY solution proposed in table I appears intermediate between the Focus and the so called 'bulk' ones, with light sfermions, which have been studied so far. One notes that while most scalars will be very heavy, outside the reach of ILC, this is not the case for the heavy Higgses which are measurable. This feature is in fact basic for the interpretation of the EGRET excess since, in the present universe, WIMPs annihilate through the CP-odd Higgs boson *A*. The rate of annihilation is reinforced by the large tan$\beta$ parameter which gives a measurable width to *A* and *H* and which, as will be shown later, may extend the mass reach of ILC.

The gaugino mass spectrum is, with the exception of the gluino, entirely accessible at ILC. One can therefore predict with very high accuracy the amount of DM. Figure 5 shows that, if real, the EGRET effect should soon also be seen in direct detection through CDMS in the Sudan mine.

**The LEP solution**

In [17,18], the excess of data reported in the Higgs search at LEP2 is interpreted as due to the production of $h(98)Z$ and $H(115)Z$, respectively. These effects are small, 2.3 and 1.7 standard deviations (s.d.), respectively. If interpreted in terms of the 2 doublet model of MSSM, they would lead to viable solutions in terms of all observables. Moreover they would provide a SUSY solution with small FT (recall that the origin of FT is the $h$ limit at 114 GeV). Finally, and of interest for the present paper, they would provide a way to control DM even for the case of a pure bino since $s$-channel annihilation is very efficient, provided that the neutralino is not too far from the Higgs pole mass divided by 2. One can therefore choose a light bino with mass ~50 GeV. The quantity $\mu$ is assumed to be large ~900 GeV as is the case in usual mSUGRA.

We will not assume that there is correlation between the sfermion and the Higgs masses and, in the spirit of this paper, one assumes $m_0$ large, say 2 TeV. Finally, to explain the low mass of $h$, one could assume the mass of the lightest stop to be ~300 GeV with an appropriate choice of the parameter $A_t$.

**The degenerate scenario within AMSB**

In AMSB, gauginos masses are $\sim \alpha\, m_{3/2}$ (loop) while squark masses go like $m_{3/2}$, with $m_{3/2}$ being the mass of the gravitino. The squarks are therefore much heavier than the gauginos, a scenario reminiscent of SpS but with only two orders of magnitude. The sleptons can have a negative contribution to their square mass but this aspect can be fixed by adding $D$ terms (avoiding any perturbation on flavour [19]).

There is a different hierarchy between the gaugino masses which leads to marked differences with respect to mSUGRA in the DM sector. One has $M_1=3M_2$ which means that the LSP is a wino. It can therefore easily annihilate into $W^+W^-$. To satisfy WMAP one needs $M_2=2$ TeV, hence a 2 TeV neutralino, a 20 TeV gluino and 800 TeV squark. Again, one needs not saturate WMAP with neutralinos and, within AMSB, gravitinos (if they can be produced during reheating) can decay into neutralinos and add a substantial component to DM.

If $M_2 \gg \mu$, the lightest neutralino is a Higgsino and to satisfy WMAP one again needs $\mu=1$ TeV. One therefore concludes that this scenario will be beyond LHC/LC unless of course the LSP does not saturate the WMAP bound.

For these reasons and also to illustrate the potential of discoveries within ILC, we have chosen a scenario where $\mu$ is small and $M_1$ and $M_2$ very large. This results in almost degenerate $\chi_1$, $\chi_2$ and $\chi_1^\pm$ and requires the ISR technique developed at LEP2 to select signals out of large SM backgrounds. An almost equivalent scenario, within ASMB, would be to assume a wino LSP, in which case only the first chargino and the LSP would be degenerate in mass.

**Summary of scenarios**

| SUSY Parameter/ Scenario | $m_\chi$ GeV | $m_{\chi_2}$ $m_{\chi_3}$ $m_{\chi_4}$ GeV | $m_{\chi_1^\pm}$ $m_{\chi_2^\pm}$ GeV | $M_1$ $M_2$ $m_{1/2}$ GeV | $\mu$ GeV | $\tan\beta$ | $m_0$ TeV | Higgs masses GeV | $\Delta\rho \times 10^4$ $g-2 \times 10^{10}$ $BR(s\gamma) \times 10^4$ CPV phases |
|---|---|---|---|---|---|---|---|---|---|
| I Focus | 378 | 430 444 739 | 417 739 | 407 724 900 | 427 | 10 | 12.5 | $m_h$=130 | 0.05, 0.02, 3 $\phi_\mu$ large |
| II SpS | 261 | 341 343 519 | 323 581 | 281 560 700 | 340 | 5 | $10^6$ | $m_h$=160 | 0, 0, 3.4 $\phi_\mu$ large |
| III h-ann. (SpS) | 79.5 | 156 410 411 | 156 416 | 78 156 201 | –400 | 5 | $10^6$ | $m_h$=169 | 0, 0, 3.4 $\phi_\mu$ large |
| IV EGRET | 64 | 116 225 250 | 115 252 | 68 128 165 | 212 | 51 | 1.4 | $m_A$=315 $m_h$=114 | 0.02, 10, 2.9 |
| V LEP | 59.6 | 115 904 | 105 900 | 60 117 151 | 900 | 20 | 2.0 | $m_h$=97 $m_H$=115 $m_A$=98 | 0.2, 1.1, 5.9 |
| VI Degen. | 299 | 301 5000 5000 | 300 5000 | 5000 5000 5000 | 300 | 20 | 5.0 | | |

**Table I**: *Solutions retained for DM studies. The main SUSY parameters are given. All but the last saturate the WMAP DM solution.*

Table I summarizes the 6 scenarios previously discussed:

- in this table $\mu$ is generated using SuSpect for mSUGRA solutions. Note however that the EGRET solution with a very large $\tan\beta$, is very dependent on the top mass. To remain compatible with reference [15] it was necessary to shift slightly the top mass to a value compatible with errors
- for the SpS solutions, one has to proceed differently given that in these scenarios $\mu$ cannot be derived from EWSB but is simply related to the gaugino masses by imposing the WMAP constraint
- Micromegas is used independently of SuSpect to allow for the SpS scenario. It gives a Higgs mass too large (with respect to SuSpect) for very heavy sfermions. This effect cannot be ignored since it has a direct influence on the amount of DM through *h*-annihilation of the neutralinos. In practice the stop mass was adjusted to generate the proper Higgs mass.

With the same SUSY parameters, one finds different gaugino masses with SuSpect and Susygen. This is probably related to a more sophisticated treatment within SuSpect (e.g. the radiative corrections to the chargino and neutralino masses have been included) but since Micromegas tends to agree with Susygen, for the present analysis we have retained the gaugino masses given by Susygen. This discrepancy illustrates the need for a consistent treatment but has little impact on the final result.

A few checks have been done:

- are these solutions compatible with the usual constraints[2] ($\rho$, $g$–2, $b{\rightarrow}s\gamma$)?
- are these solutions consistent with the limits given by Direct and Indirect searches?
- for the EDM, we have chosen a Focus solution which allows a large phase for $\mu$ given the present limit on the electron EDM. For SpS solutions this condition is naturally satisfied.

All solutions pass these tests when they are meaningful (for SpS only the CPV feature is displayed) except for the LEP solution for what concerns $b{\rightarrow}s\gamma$ which comes out too large. It has been argued [11] however that this kind of test is not necessarily meaningful since the SUSY flavour sector is poorly controlled. The $g$–2 value determined in $e^+e^-$ does not favour heavy sfermions and only the EGRET solution improves the agreement. All but one (the degenerate case) saturate the WMAP DM value. Some of these solutions should be testable soon either indirectly by DM searches (EGRET) or by EDM measurements (SpS).

## III Experimental issues

In this section, we will describe the strategy needed to separate SUSY signals from the SM backgrounds. Without going into the details, we will show that although backgrounds are 100 to 1000 time larger than some signals, it is possible, at ILC, to devise robust selections which provide pure signals with reasonable efficiencies. This has already been done in some detail for the charginos with similar signal/background ratio in [21]. For what concerns neutralinos, the present scenario is much more challenging given that, with heavy sfermions, the cross sections are smaller than usually assumed and needs new selections based on b-tagging.

Finally, for what concerns the degenerate case, the use of ISR will be evaluated and we will show that photon-photon interactions, in spite of their extremely high cross section, do not contribute appreciably.

**SM backgrounds**

Figure 6 shows that the SM backgrounds are huge as compared to some expected signals.

The $We\nu$ channel, with a cross section of several 1000fb, poses a severe problem to detect the channel $\chi_1\chi_2$ often in the few fb range. The neutralino $\chi_2$ decays into $\chi_1$ and a $Z$, most often virtual. Only the leptonic and $b$ quark channels, the latter with extreme purity and efficiency, can be separated from this type of background.

The $ZZ$ channel with one $Z$ into neutrinos and the other into $b$ quarks, can partially be eliminated since it tends to give forward peaked $Z$ and a missing mass centred at $M_Z$. Events with ISR can still pass these cuts and therefore one needs a more detailed estimate.

---

[2] Reminder [20]: $\rho$=1.000±0.001, $\delta a_\mu(e^+e^-)$=(24±10)×10$^{-10}$, $\delta a_\mu(\tau^+\tau^-)$=(7.6±9.0)×10$^{-10}$ and $BR(s\gamma)$=(3.25±0.37) ×10$^{-4}$.

Since the neutralino channels go into $Z^*$, one could naively expect some rejection using mass reconstruction and eliminating on-shell $Z$ (or $W$). Unfortunately this method only allows a modest rejection, given the natural spread of the Breit-Wigner mass distribution.

The only promising method is therefore to select pairs of bottom quarks which would essentially eliminate the $WW$ and $We\nu$ channels since $W$ decays into $b$ and $c$ in one per mill of the cases (one can also control this elimination on the basis of the reconstructed mass which would peak at the $W$ mass).

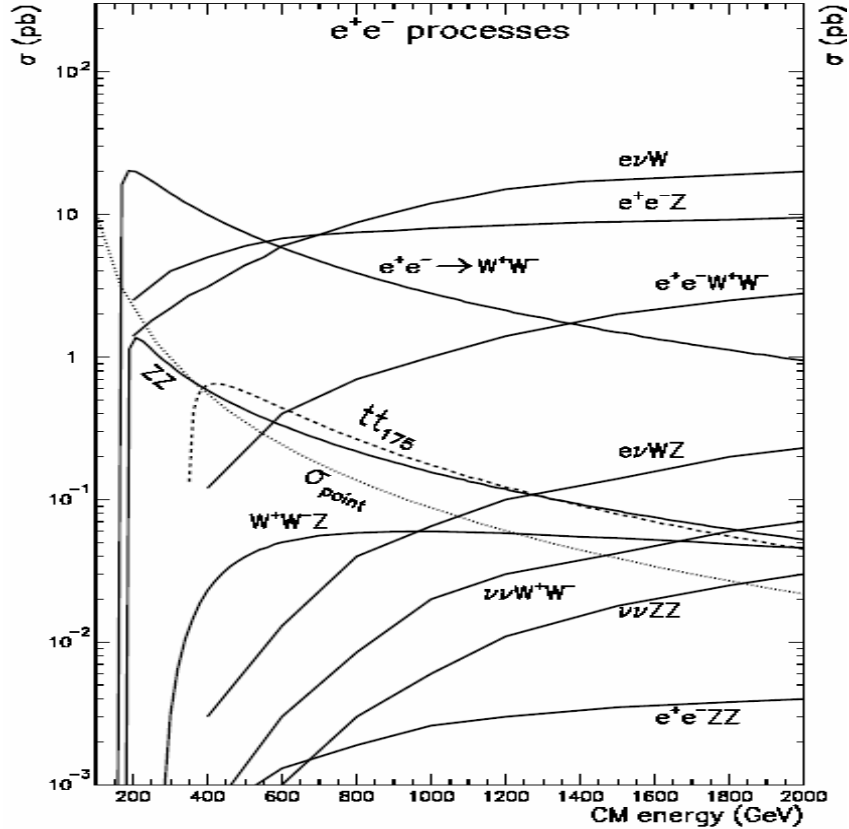

How much rejection can be expected? A factor 1000 would be ideally needed which is certainly very challenging but recall that this rejection is against $c$ and $s$ and not against $c$ and $c$ as for a $Z$. One also needs a good efficiency given the low cross section. From [22] one expects an efficiency above 50% for this type of rejection.

The contamination from $Zee$ can be removed by requesting a $Z^*$ with large missing transverse momentum and no isolated electron.

**Figure 6:** *SM process cross sections versus the centre of mass energy [23].*

From this qualitative analysis, one concludes that for neutralino channels, like $\chi_1\chi_2$ and $\chi_1\chi_3$, the most challenging background comes from $ZZ$.

**Charginos**

From table II, one sees that chargino cross sections are in the 100-1000 fb range. The background comes from $WW$, $We\nu$, $ZWW$ (10 fb assuming that $Z$ decays into neutrinos) and $WW\nu\nu$ (few fb). The two latter backgrounds are not obviously reducible. One can presumably gain a factor 10 by requesting an off-shell $W$ (when at least one $W$ decays hadronically). These two channels should therefore contribute below 1fb.

Much more of a concern is $WW$ with its huge cross section. This process is reducible since it has a very specific topology, forward peaked and back-to-back $W$'s. The latter property remains true in the transverse plane even when ISR is taken into account (clearly ISR at angle has to be vetoed). Can one reach a rejection better than a thousand on this channel? The answer to this question clearly requires a detailed study with fully realistic generators and detector response.

Why is this rejection so critical after all? The reason, as will be discussed below, is the measurement of the charge asymmetry in the chargino channel which is needed to achieve optimal extraction of SUSY parameters. Any remaining contamination of the *WW* component is dangerous since its charge asymmetry is much stronger than for the signal.

Using a simplified simulation [24], and standard selections based on missing transverse momentum and a veto on forward energy, it can be shown that these problems are manageable since the background, as seen in figure 7, can be reduced to a very low level. This result assumes that the detector PFLOW (particle flow) properties are optimal, giving a hadronic mass resolution of order ~30%/$\sqrt{M}$. Figure 7 also shows that the right hand side of the mass distribution has a clean edge thus allowing a precise determination of the mass difference between the chargino and the neutralino masses.

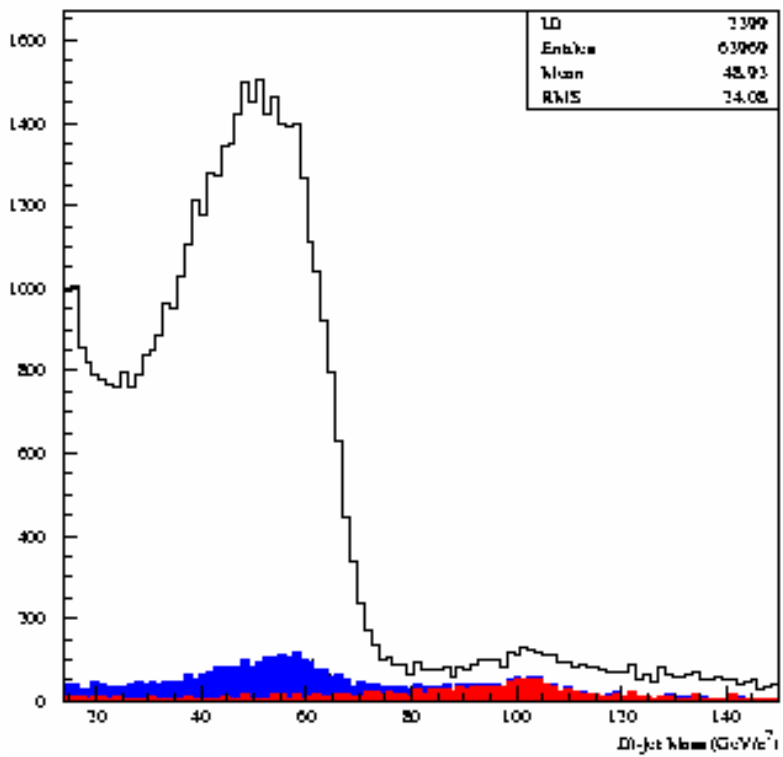

One may also worry about the definition of a charge asymmetry given the incomplete reconstruction of the chargino channels. To define the charge one has to request that one of the $W^*$ decays leptonically which only adds to incompleteness. We believe however that this problem can be tackled with appropriate algorithms.

**Figure 7:** *hadronic mass distribution from semi-leptonic decay events in the chargino analysis in the h-annihilation scenario at 500 GeV. The blue (low masses) histogram is from WW while the red one is from We ν.*

Charginos are produced through *s*-channel annihilation into a combination of *Z* and *γ* and through sneutrino exchange. Polarisation of the initial electrons is an essential tool:

- to measure the chargino mixing in terms of winos and Higgsinos
- to estimate the sneutrino contribution.

Sneutrino exchange can be decoupled by using right-handed polarisation. It will be shown (Focus scenario) that combined with charge asymmetry, the right-handed cross sections allows to measure the wino-Higgsino composition. The mixing angles then serve to deduce precisely the SUSY mass parameters and tan$\beta$ (less precisely) even if the heavier chargino is kinematically inaccessible. Knowing the wino-Higgsino composition, one can then use left handed polarisation to measure the sneutrino exchange contribution. It turns out that this measurement is very sensitive through interference with the *s*-channel term, well beyond the ILC/LHC direct reach.

| Channel: Masses Decay modes $\sigma$(ECM) fb $\sigma_{bkg}$ fb | ECM GeV | $\chi_1^+\chi_1^-$ | $\chi_1\chi_2$ | $\chi_1\chi_3$ | $\chi_2\chi_3$ | $\chi_1^+\chi_2^-$ or $\chi_2^+\chi_2^+$ | Specific channels |
|---|---|---|---|---|---|---|---|
| I Focus | 950 | 417×2 $W^*W^*$ 89×0.35 | 378+430 $Z^*$ 17×0.10 0.05 | 378+444 $W^+\chi^-$ <1 | 430+444 $Z^*W^*W$ 16×0.50 0.5 | 417+739 | |
| II SpS | 750 + 1000 | 323×2 $W^*W^*$ 164×0.35 | 261+341 $Z^*$ <1 | 261+343 $Z^*$ 17×0.10 0.05 | 341+343 $Z^*Z^*$ 33×0.30 0.05 | 323+581 $Z$(47%)$W^*$ 4×0.05 0.2 | $\chi_1\chi_4$ 261+519 67% $h\chi_2$ 4×0.5 0.05 |
| III h-ann. | 650 + 900 | 156×2 $W^*W^*$ 500×0.4 30 | 79+156 $Z^*$ <1 | 79+410 0.1$Z$ 0.7$W$ 1 | 156+410 $Z^*WW$ 3×0.5 0.5 | 416×2 0.16$W\chi$0.41$Z\chi^{\pm}$ 0.43$W\chi_2$ 78×0.6 0.1 | $\chi_3\chi_4$ 410+411 0.72$W\chi^{\pm}$ 0.21$Z\chi_2$ 32×0.3 0.5 |
| IV EGRET | 500 | 115×2 $W^*W^*$ 684×0.3 30 | 64+116 $Z^*$ 5.2×0.1 0.05 | 64+225 0.1$Z$ 0.9$W$ 21×0.01 0.05 | 116+225 $Z^*WW$ 38×0.5 0.5 | 252×2 0.4$W\chi$ 0.6$Z\chi^{\pm}$ 334×0.15 0.05 | $\chi_3\chi_4$ 225+250 0.85$W\chi$ 0.1$h\chi$ 100×0.05 0.05 |
| V LEP | 300 | 105×2 $W^*W^*$ 2300×0.4 20 | 60+105 <1 | <1 | <1 | 900×2 | *H, A, h* Accessible |
| VI Degen. | 800 | 300×2 150×0.02 1 | 299+300.7 63×0.02 1 | | | | ISR + 2 prongs |

**Table II:** *For each of the six scenarios defined in the text are indicated the centre of mass energy (ECM) assumed (second column), the dominant final states (with BR), the signal cross section×efficiency and the remaining background (bkg) cross section after selections.*

An estimate of backgrounds, obtained with the simplified simulation [24], is given in table II for the various scenarios. It is based on a straightforward selection approach and therefore can certainly be improved using optimized approaches as for LEP2 analyses. In the following we will not take into account this limitation of performances and assume ideal performances. Note also that in some cases one is using right-handed polarization which therefore essentially eliminates the *WW* background contribution.

**Neutralinos**

Table II shows that, with heavy selectrons, some neutralino signals to be studied can be at the level of 1-10 fb with huge SM backgrounds. Cross sections with identical neutralinos have not been shown since they tend to be even smaller: due to Fermi statistics, they are produced into a *p*-wave and therefore suppressed at threshold. This is also the case for neutralinos with same CP values. In some cases it affects $\chi_1\chi_2$ in others $\chi_1\chi_3$. The $\chi_2\chi_3$ channel is in contrast always unaffected since it comprises CP opposite neutralinos.

As explained in the preceding section, neutralinos are selected using *b*-tagging. This selection also takes care of removing the chargino contamination since charginos go into *W* modes.

For $\chi_1\chi_2$ and $\chi_1\chi_3$ channels into $\chi_1\chi_1 Z^*$, the main contaminant is $ZZ \rightarrow bb\nu\nu$ with ~25 fb cross section. Then:
- removing events with $|\cos\theta_Z|>0.9$ leaves us with ~3fb, *ZZ* being forward-backward peaked since it proceeds through electron exchange
- vetoing against on shell *Z* (reconstructed mass and recoil mass) retains 0.4 fb with no effect on the signal which goes into $Z^*$.

The signal efficiency is about 10% if one retains lepton pairs and *b* quark final states. The remaining background corresponds to *ZZ* events with an ISR photon. Given that in most cases the lightest neutralino is heavier than 100 GeV, one can require a missing mass heavier than 200 GeV which implies an energetic ISR photon and therefore a further suppression of about 10 of the background.

This strategy does not apply for the *h*-annihilation and EGRET solutions for which the dominant decay mode of $\chi_3$ is $W^+\chi_1^-$ which will therefore be overwhelmed by the chargino channel.

| Neutralino $\Delta m$ in GeV | $\chi_1$ | $\chi_1\chi_2$ | $\chi_1\chi_3$ | $\chi_2\chi_3$ | $\chi_2^\pm$ | Specific |
|---|---|---|---|---|---|---|
| Focus | 0.5 | 0.3 | – | 0.1 | – | – |
| SpS | 0.4 | – | 0.3 | 0.1 | – | $\chi_1\chi_4$ 0.3 |
| h-ann. | 0.2 | – | – | – | 0.1 | $\chi_3\chi_4$ 0.1 |
| EGRET | 0.2 | 3 | – | 0.1 | 0.1 | $\chi_3\chi_4$ 0.2 |
| LEP | 0.2 | – | – | – | – | – |

**Table III:** *Expected accuracies on the neutralino mass determination using a scan method with an integrated luminosity of 100 fb$^{-1}$ near threshold combined with 500 fb$^{-1}$ at an energy defined in table II.*

For $\chi_2\chi_3$ into $\chi_1\chi_1 Z^*Z^*$, the contaminant is again *ZZ*. Similar cuts can be applied with a similar rejection and efficiency ~30%. When $\chi_3$ decays into a chargino and a *W*, the final state is $Z^*W^*W$ with a large missing mass. The standard cross section given in figure 7 is ~50fb. A reduction by 10 is obtained requesting an energetic ISR photon and one can also perform other cuts depending on the topology. For purely hadronic decays two ISR photons are needed to have a large missing mass. One can also require that the gauge bosons are not on mass shell. With this selection one expects 50% efficiency and a 0.5 fb background which is sufficient for this channel with a large cross section.

Table III gives the accuracy on neutralino masses expected from the scan method described in Appendix II. Note that the LSP mass is usually measured from Z leptonic decays from the neutralino modes. With the present scenarios this will not be, in general, possible. One can instead use the mass distribution of $W^*$ hadronic decays of charginos. The end point of this mass distribution will give the mass difference between the chargino and the LSP (see figure 7). The former can be accurately determined with a scan while the mass distribution, with high statistics and excellent hadronic mass resolution (<5%) provided by ILC detectors, will be essentially dominated by systematics. Figures given in table III are only estimates.

The luminosity budget for this analysis is the following. One assumes that 500fb$^{-1}$ are spent on the chargino channels (1000fb$^{-1}$ in case this channel is produced above 500 GeV). This luminosity is also used to produce nearby neutralino channels. Scans on neutralinos, if relevant, are based on an extra 100fb$^{-1}$ per channel near threshold. Accuracies on cross sections can simply be deduced from table II given the rates assumed.

**'Chargino counting'**

This technique is of potential use in a variety of SUSY scenarios. Three examples are:

- chargino/neutralino mass degeneracy as in scenario VI
- stau/neutralino degeneracy in the co-annihilation scenario
- stop/neutralino degeneracy in other types of co-annihilation scenarios.

While the first and third cases correspond to relatively large cross sections, stau channels are very challenging. The main background for this method is due to $\gamma\gamma$ physics with $\tau\tau$ final states plus ISR. There is considerable suppression of this background given that the final state electron (positron) is forced to remain in the beam pipe, with an angle below 5mrad. This amounts to say that this electron cannot appreciably deviate from its initial trajectory and therefore does not radiate photons. The ISR is almost entirely suppressed in γγ as compared to the annihilation processes as shown, in a simplified way, in Appendix I but confirmed using the generator BDKRC [25].

A $\gamma\gamma$ event accidentally superimposed to a 'neutrino counting' event, that is $e^+e^- \rightarrow \nu\nu\gamma$, can also fake this type of event. Selecting the most dangerous topologies, that is $\gamma\gamma \rightarrow \pi^+\pi^-$ and $\gamma\gamma \rightarrow \mu^+\mu^-$, one finds that there will be 0.15 events per beam crossing giving the right topology. These events can be rejected by requesting that the pions (or muons) are unbalanced in transverse momentum but about half of these events will come from $\gamma\gamma^*$ and therefore will have a spectator electron (positron) carrying a finite transverse momentum.

On top of neutrino counting, another potential source of 'fake' neutrino counting events is $e^+e^- \rightarrow e^+e^-\gamma$ when $e^+$(or $e^-$) is missed by the forward veto. This may seem to be a dangerous background given the large cross section of the diffusion process at very forward angles but recall the previous argument on radiation suppression which also applies in this case and therefore the process poses no particular problem.

# IV Determination of SUSY parameters

In this section, we indicate the methods used to extract the relevant quantities needed to evaluate the DM contribution of SUSY in the scenarios under consideration. We will show that, in spite of an incomplete coverage of the SUSY spectrum, ILC is able to recover the missing information.

Beam polarisation will play an essential part in this strategy and we recall that the effective beam polarisation can be precisely reconstructed using the *WW* data as was suggested in [26].

**How to recover 'missing' parameters**

It should be first noted that one needs to prove that the SUSY scalar fermions are indeed heavy and not only the squarks as can be proven from LHC data but also sleptons, in particular selectrons, which can contribute to the gaugino cross sections and therefore to the processes influencing the amount of DM. To prove this, one would search for these sleptons at LHC in the gluino cascades but this is not possible in the set of scenarios under consideration since in all cases the gluino is lighter than the sleptons. LHC would therefore only set a poor lower limit on slepton masses.

We will show how the sneutrino mass can be indirectly estimated using the chargino channel. For a 1 TeV sneutrino, this mass can be measured at the per cent level, while the sensitivity for very high masses reaches 10 TeV, decreasing for a large Higgsino component. These results are largely sufficient to prove the 'heavy scalar' scenario and to reduce to a negligible level the effect of these unseen scalars, avoiding model dependent assumptions.

In some cases one can only have a partial coverage of the gaugino sector and therefore not be able to extract the SUSY parameters relevant for DM in a fully model independent way. This can happen when $M_2$ or $\mu$ are above 500 GeV which means that the heavier charginos and the fourth neutralino $\chi_4$ cannot be observed at ILC. One can assume a GUT relation between $M_1$, $M_2$ and $M_3$ (and even check it for $M_1$ and $M_3$ by measuring the gluino mass at LHC), but this approach is certainly not fully satisfactory.

We will show how the high accuracy measurements with polarized beams performed with the lightest chargino can help solving this problem.

**The Focus solution**

This solution illustrates the issue of determining indirectly some SUSY parameters in case of a Higgsino-like chargino. The second chargino is not accessible and therefore the analysis will rely solely on the lightest chargino. We know from the chargino cross section itself that the lightest chargino is dominantly a Higgsino. This can be proven accurately by measuring the right-handed cross section which should be maximum:

$$R = \sigma_R / \sigma_R^{max} = \left[ (1-c_{2L})^2 + (1-c_{2R})^2 \right] / 8 \quad \text{with } c_{2L,2R} \equiv \cos 2\phi_{L,R}.$$

One has $c_{2L}$ and $c_{2R}$ close to $-1$: $c_{2L} = -0.953$ and $c_{2R} = -0.885$ and $s_{2L} = -0.30$, $s_{2R} = -0.47$ (always negative).

The following steps are taken:

1/ One estimates $\mu$ from the second neutralino mass (this mass being relatively insensitive to other parameters).

2/ The cross section $\sigma_R$~30 fb (at 950 GeV) is known with a statistical error of 0.6% for 1 ab$^{-1}$.

3/ Then one uses the right-handed charge asymmetry $(A_{FB})_R \sim 3(c_{2L}-c_{2R})/8$ to obtain $c_{2L}-c_{2R}=-0.07\pm0.02$.

From [27]: $M_2-\mu=\sqrt{2}M_W q(\sin\beta+\cos\beta)$ where $q=\pm(c_{2L}+c_{2R})/(s_{2L}+s_{2R})$ is measured with a 7% relative error which comes mainly from the charge asymmetry measurement. One finds two solutions $M_2$=427±274 GeV for large tan$\beta$ leading to $M_2$=701 GeV (the minus-sign solution is incompatible with observation). For tan$\beta$=1, one has $M_2$=814 GeV.

One can then use $c_{2L}-c_{2R} \sim 4M_W^2 \cos 2\beta/(M_2^2-\mu^2)$ which clearly allows to exclude tan$\beta$=1. This relation gives cos2$\beta$=−1.00±0.3 and therefore tan$\beta$>2.4. With this value it is possible to say that cos$\beta$+sin$\beta$<1.31 hence $M_2$< 785 GeV. Taking into account the error due to $q$ one finds that $M_2$ can be determined to 5.7%.

4/ The sneutrino exchange only affects the left-handed cross section. For a pure Higgsino, there is no sensitivity to sneutrino exchange and therefore one only expects a rather weak sensitivity. The statistical accuracy expected at ILC on $\sigma_L$ is 0.24% at 950 GeV with 1 ab$^{-1}$. If the polarisation uncertainty can be maintained below this value, the sensitivity could be pushed up to 3 TeV which is still insufficient to reach the true value at 12.5 TeV but is largely sufficient to establish the Focus scenario and remove the corresponding DM uncertainty.

**The SpS solution**

This scenario will be identified by a $h$ heavier than 150 GeV (not only due to large $m_0$ but also to the running of $\lambda$ in the Higgs potential).

One can in principle measure the second chargino mass using the $\chi_1^+\chi_2^-$ channel but this channel will not yield a precise number given its low cross section and efficiency. The neutralino channels can be used instead since, as shown in table III, all neutralino masses can be measured accurately. In particular note that this is true for $\chi_1\chi_4$ since $\chi_4$ decays into $h\chi_1$. These features allow to conclude that tan$\beta$=5±1.

**$h$-annihilation within SpS**

This solution seems ideal for ILC since the heavier charginos will be accessible and therefore one reaches optimal DM measurements. ECM=900 GeV is needed to produce the second chargino.

The quantity tan$\beta$ can be determined using the FB asymmetry and the second chargino mass information through the formula [27]:

$$\tan\beta = \sqrt{\frac{4M_W^2 - \left(m_{\chi_2^\pm}^2 - m_{\chi_1^\pm}^2\right)(c_{2L} - c_{2R})}{4M_W^2 + \left(m_{\chi_2^\pm}^2 - m_{\chi_1^\pm}^2\right)(c_{2L} - c_{2R})}}$$

The main uncertainty on $\tan\beta = 5.0 \pm 0.8$ is due to the error on the mixing parameters measured with the charge asymmetry.

**EGRET solution**

In the EGRET case all gauginos are light which makes it a show case for ILC. The sensitivity to the sneutrino mass is excellent allowing, through the chargino channel, to determine this mass at the 1% level or, alternatively, to be sensitive to sneutrino masses up to 12 TeV.

One can access to the heavier Higgses through the *HA* channel. There is a simple signature for this channel which primarily gives 4*b* quarks. For *HA* into 4*b*, the main background comes from the SM *bb* channel with radiation of a virtual gluon decaying into *bb*. From LEP2 data, one can estimate that this background should be below 0.1fb. The four fermion background, *ZZ* is negligible. The width of *H/A* is ~30 GeV and therefore can be measured. Since it goes like $\tan^2\beta$, this quantity can be precisely determined. Figure 8 shows the effect of the width on the *HA* cross section which is enhanced at threshold. This allows extending significantly the mass reach of ILC for this channel.

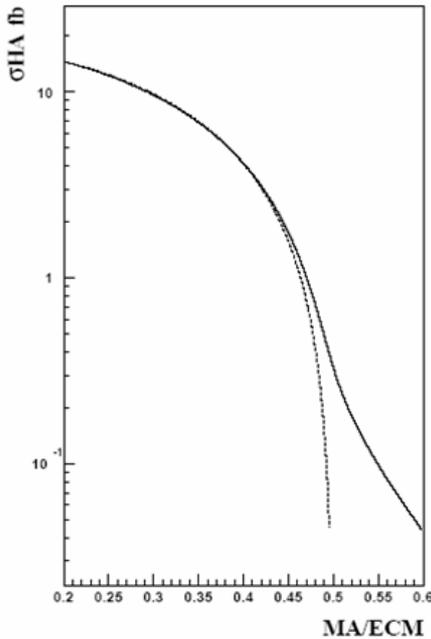

**Figure 8:** *The HA cross section for the EGRET solution, at fixed CM energy versus the heavy Higgs mass normalized to the centre of mass energy. This plot shows that the large width of the resonances allows producing them above the kinematical limit.*

**LEP solution**

This solution illustrates the capability of ILC to determine indirectly the $\mu$ parameter in a scenario where the lightest chargino is a gaugino. One cannot therefore proceed as for the Focus case since the right-handed cross section vanishes.

The following procedure is applied:

- from the Higgs sector one determines $\tan\beta$,
- from the energy dependence of the lightest chargino cross section one extracts the sneutrino mass, $2.0 \pm 0.1$ TeV, which allows to subtract the sneutrino contribution for the chargino analysis
- from the chargino charge asymmetry and cross section one computes the combination *pq* defined in [27]
- $M_2$, known from the chargino mass, gives a second relation in terms of *p*, *q* and $\tan\beta$, which allows to compute $\mu$.

In the non-decoupling regime corresponding to this scenario, one can easily measure $\varepsilon=\beta-\alpha$ using the $Zh$ channel which gives $\sin^2(\beta-\alpha)\sim 0.1$ at the per cent level. For a large $\tan\beta$, the three MSSM Higgs bosons will decay almost exclusively into $b$ quarks and $\tau$ lepton, which forbids an accurate determination of the $\beta$ parameter itself. Recently, it was suggested to determine $\tan\beta$ using $\tau\tau$ fusion in the $\gamma\gamma$ mode [28]. This would allow a precise determination of $\tan\beta$ which considerably improves the indirect determination of $\mu$.

The cross section of the fusion process $\gamma\gamma\rightarrow h\tau\tau$ is large since the coupling goes like: $\sin\alpha/\cos\beta=\tan\beta\cos\varepsilon-\sin\varepsilon\sim\tan\beta$. This formula shows that for the non-decoupling case $h$ provides a precise determination of $\tan\beta$. Note also that in the decoupling case ($\varepsilon=\pi/2$), $h$ has, as expected, a standard coupling independent of $\tan\beta$.

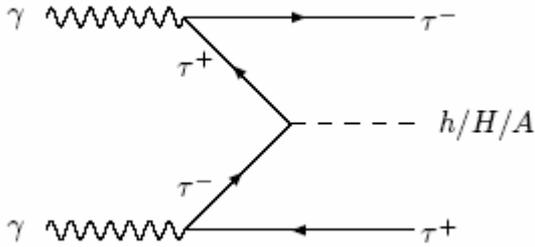

One can estimate that with $\tan\beta=20$, the cross section for this channel is ~5 fb for a centre of mass energy of 400 GeV. Using the $bb\tau\tau$ final state one could, optimistically, reconstruct a clean signal and deduce the coupling at the 5% level for 200fb$^{-1}$. One can therefore hope to determine $\tan^2\beta$ at the same level of accuracy.

It should also be noted [29] however that other approaches are possible based on $e^+e^-\rightarrow bbh, bbA, bbH$ which combined with width a measurements based on $e^+e^-\rightarrow hA, HA, H^+H^-$ can also allow to extract $\tan\beta$.

Knowing $\tan\beta$, one then follows a strategy as for the Focus scenario but, in the present case, it is not possible to use the right-handed cross section (vanishing in the gaugino case) to remove the effect of sneutrino exchange. Instead one will determine the sneutrino term from a study of the energy dependence of the cross section. Subtracting this term one can then follow the procedure defined previously. From the cross section and the charge asymmetry measurements one gets $pq=-63\pm14$ where the error is due to the latter. The quantity $M_2$ is estimated from the chargino mass and gives a second relation defining $p$ and $q$. Solving these relations, one finds $\mu=900\pm100$ GeV.

In conclusion the LEP solution tells us that one can measure $\mu$ with a precision of 10 % assuming that $\tan\beta$ can be provided by the $\gamma\gamma$ collider mode of ILC.

**Mass degeneracy**

In this scenario one uses the ISR method already developed at LEP2 which, with modest luminosity, has allowed covering the whole mass spectrum. In Appendix I are recalled the elementary formulae which describe ISR for the annihilation process and for the $\gamma\gamma$ background, the latter being negligible.

In the presently considered scenario one can use this method to access to the lightest chargino and to the second lightest neutralino channels. From [30] one finds that, for a mass difference of 800 MeV, the chargino channel decays into one prong events which are made of 60% $\pi^{\pm}$ and 40% $e\nu$ and $\mu\nu$. The $\pi\chi$ is a two-body decay mode which allows to directly extract the mass difference Δ*m* between the chargino and the neutralino. Moving to the centre of mass system *S* of the two charginos, which is defined knowing the energy of the ISR photon, one can plot the energy difference between the two pions divided by $\beta\gamma$ ($\beta\gamma = p/m$ where *p* is the momentum of the chargino in *S* and *m* is the chargino mass) in the rest frame of chargino and it is easy to prove that this quantity, centred at zero, is distributed as a triangle with a basis equal to 4Δ*m*.

The following selections were applied:

- 2.5°<$\theta_\gamma$<177.5°
- 35 GeV<$E_\gamma$<175 GeV
- 2 total energy of the two charged particles below 3 GeV
- total transverse energy greater than 2.75 GeV.

These leaves about 1000 events for the signal. The $\gamma\gamma$ background was added randomly to neutrino counting candidates. Electrons were eliminated assuming d*E*/d*x* identification with the TPC. These leaves about 850 background events. One can easily reduce this background with a more sophisticated analysis left for further developments. The neutrino counting cross section can further be reduced by an order of magnitude using right-handed electrons while keeping most of the chargino cross section in the case of a Higgsino-like solution.

Assuming no background, one finds that Δ*m* can be known to ~15 MeV. We have assumed an error of 20 MeV. Due to co-annihilation, this error dominates the contribution for DM.

The chargino mass can be estimated by adjusting the end point of the photon spectrum shape indicated in figure 9. This type of determination can give the chargino mass at the per cent level. Knowing the cross section at the per cent level, one can establish that the chargino is a pure Higgsino. The chargino mass measurement provides *μ* with 1 GeV accuracy.

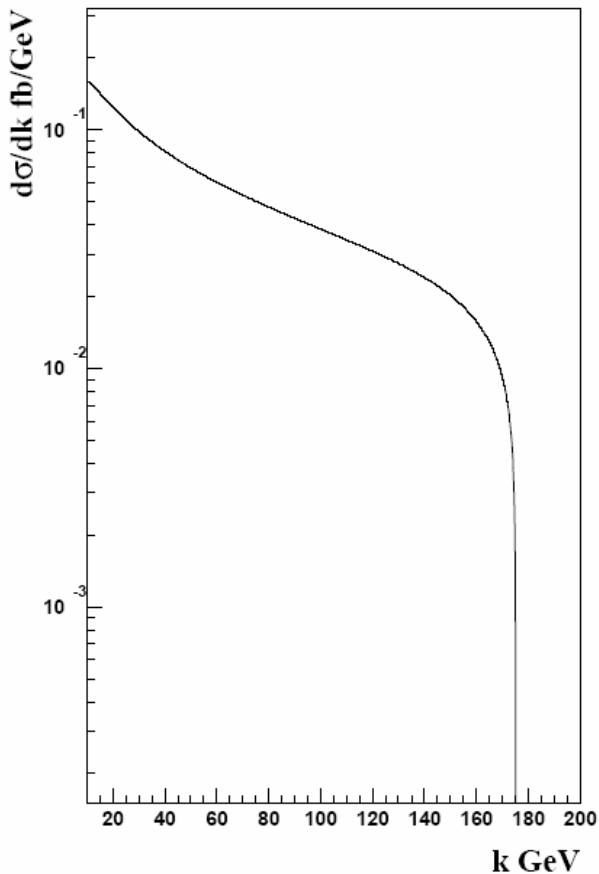

From these expressions one sees that $M_1$, $M_2$ and sin2$\beta$ vary over large intervals but in a correlated way:

- tan$\beta$~35  $M_2$=4 TeV   $M_1$~ ∞
- tan$\beta$=1  $M_2$=7.5 TeV $M_1$=2.2 TeV.

These uncertainties, as was checked with Micromegas, are not critical for the determination of DM.

One finds $\Omega h^2$=(1.00±0.01)%.

**Figure 9:** *This figure displays the ISR spectrum expected in the degenerate scenario in fb/GeV at a centre of mass energy of 800 GeV.*

From the measurement of Δ*m*, one can infer that the other gauginos are very heavy. Using the formalism of [27], one has two approximate equations, valid for $M_2, M_1 \gg \mu$:

$$\Delta m' = m_{\chi_2} - m_{\chi_1} = M_Z^2 \left( \cos^2 \theta_W / M_2 + \sin^2 \theta_W / M_1 \right)$$

$$\Delta m = m_{\chi_1^\pm} - m_{\chi_1} = M_Z^2 \left( \sin^2 \theta_W / M_1 - \cos^2 \theta_W / M_2 \right) \sin 2\beta + \Delta m' / 2$$

where $\Delta m'$ and $\Delta m$ can be very precisely determined and therefore set the strongest constraints.

**Summary**

In case of an incomplete coverage by ILC of the SUSY mass spectrum relevant to DM (heavy sleptons, heavy Higgses, heavy gauginos), the key to access to the SUSY parameters is the lightest chargino channel using polarisation and charge asymmetry. In case the heaviest chargino is not accessible at ILC, it can provide an estimate of $\mu$ or $M_2$ and $\tan\beta$.

The lightest chargino can give an impressive indirect sensitivity on the sneutrino mass as shown in figure 10. This sensitivity is reduced for a Higgsiso-like chargino (small $\mu$) but, in this case, the slepton exchange plays no role in the DM prediction.

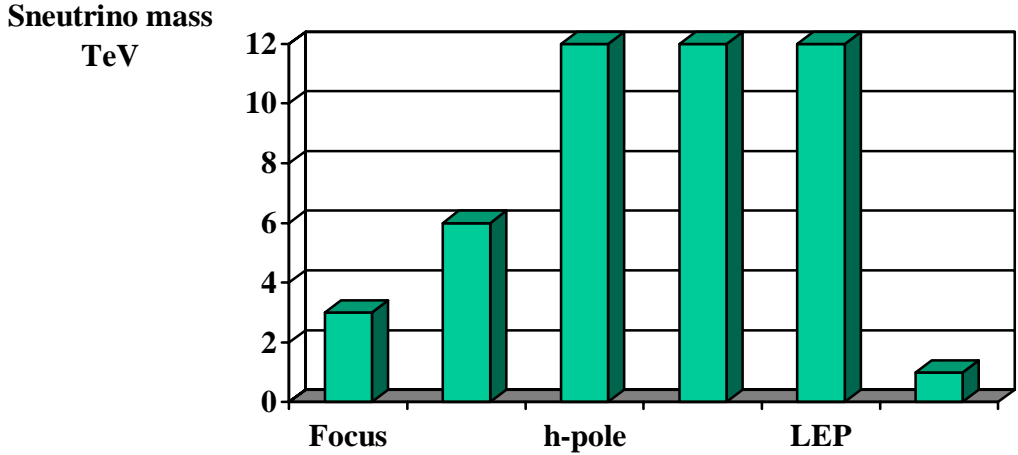

**Figure 10:** *Indirect sensitivity of ILC to heavy sneutrinos in the various scenarios described in this paper (these masses are excluded at 90% CL for ECM=1 TeV and 1ab$^{-1}$).*

## V Precision on DM

In this section we describe the methods used to estimate the precision on DM given the various observables accessible at ILC. These observables are:

- the LSP mass known at the 0.2–0.5 GeV level
- the chargino masses, cross sections with polarisation and FB asymmetries
- masses (~0.5 GeV level) and cross sections (~1% level) for some heavy neutralinos
- sneutrino mass or at least a lower limit on this mass
- Higgs masses.

With these observables one determines $\mu$, $M_1$, $M_2$, $\tan\beta$ as explained in the preceding section.

The sensitivity to DM of these parameters depends very much on the type of scenario under consideration. In practical cases the amount of DM depends critically on very few parameters. After identifying these parameters, using Micromegas, they are varied within errors taking into account correlations.

The most difficult scenarios are those for which light neutralino annihilation takes place through the lightest Higgs like scenarios III and V. Why is so?

Firstly, there is a critical dependence on the LSP mass since annihilation takes place through a $p$-wave ($h$ being a CP even state). Micromegas shows that a 100 MeV error on the LSP mass can give a 5% uncertainty on DM.

Secondly, the coupling of neutralinos to $h$ critically depends on mixing angles since the couplings depend on the amount of Higgsino and wino components. The error on mixing can depend very critically on tan$\beta$, usually very poorly known unless one has access to the heavy Higgs sector. This is the limiting factor in scenario III in spite of the full power of the chargino precision measurements.

Mixing also depends on the precision on $\mu$ which, in the case of scenario V, is poorly known. Higgs production, at a $\gamma\gamma$ collider, can help in determining tan$\beta$ and therefore $\mu$ but the error on DM remains unsatisfactory.

In scenarios I and II where the charginos are Higgsino dominated, the precision on DM is excellent since one can precisely determine $\mu$ and the uncertainty on $M_2$ does not play a major role. The most demanding analysis is clearly on tan$\beta$ where one needs to fully exploit the lightest chargino precision measurements with polarized beams to reach the proper level. In scenario II one benefits from the very complete determination of the neutralino masses which add extra constraints.

The EGRET scenario is clearly ideal in the sense that the heavier charginos are accessible and that the heavy Higgs sector provides a precise determination of tan$\beta$. A $\gamma\gamma$ collider may allow to access to the heavy Higgs sector at a reduced energy.

For what concerns the degenerate scenario, the DM precision is primarily due to co-annihilation between the LSP and the lightest chargino (and neutralino) and therefore depends on mass differences between these particles which are very well determined as explained in the previous section.

Table IV and figure 11 summarize our results.

## VI Summary and conclusions

The SUSY DM scenarios analysed in this paper illustrate various capabilities of ILC. They were selected on the basis of theoretical criteria (Focus, SpS scenarios) and on some experimental indications of new physics (EGRET, LEP) related to SUSY.

These results, shown in table IV, are quite contrasted and accuracies vary over a wide range depending on the type of scenario. In the Focus case, where there is no precise determination of tan$\beta$ and $M_2$ since the second chargino is beyond reach, the DM precision is excellent. At the other extreme, for the Higgs-annihilation scenario where the two charginos are accessible, the sensitivity to tan$\beta$ is so critical that one cannot achieve an adequate accuracy. In the LEP scenario where one has assumed that tan$\beta$ could be determined from the Higgs sector, the indirect determination of $\mu$ (again the second chargino is beyond reach) is not precise enough to provide the right accuracy.

In spite of these limitations, we have found that polarisation and charge asymmetry measurements with the lightest chargino allow estimating accurately the missing gaugino parameters ($\mu$ or $M_2$), therefore avoiding a model dependent analysis. It was also emphasized that a precise study of the contamination of the WW channel is needed to estimate the contamination to the chargino channel since this background is potentially dangerous for what concerns charge asymmetry measurements. First indications, with a fast simulation, are reassuring.

These measurements also provide a very high sensitivity on the sneutrino mass. This sensitivity depends on the Higgsino-wino relative content but, as often emphasized in this paper, when sfermions are heavy, these two components are present in order to provide the right amount of DM. It is therefore not too surprising that ILC can access to sfermion masses well beyond the reach of LHC. This virtue is of course reminiscent of what has been observed in the Z' sector where similar mass reaches have been reported [31]. Figure 10 summarizes these results.

| SUSY Parameter/ Scenarios | LSP GeV | $M_2$ GeV | $M_1$ GeV | $\mu$ GeV | tan$\beta$ | Sneutrino mass or bound TeV | Features | Overall effect $\Omega h^2_{DM}$ (origin) |
|---|---|---|---|---|---|---|---|---|
| I Focus | 378±0.5 | 724±40 | 407±3 | 427±2 | 10 >2.4 | 12.5 >3 | $m_h$=130 GeV | 1% |
| II SpS1 | 261±0.4 | 560±1 | 281±1 | 340±1 | 5±1 | $10^6$ >6 | $m_h$=160 GeV | 1% |
| III h-ann | 79.5±0.2 | 156±1 | 78±1 | −400±1 | 5±0.8 | $10^6$ >12 | $m_h$=163 GeV | 40% (tan$\beta$) |
| IV EGRET | 64± 0.2 | 128±1.9 | 68±0.2 | 212±2.5 | 51 48<tan$\beta$<54 | 1.4±0.014 >12 | HA accessible | 2% |
| V LEP | 59.6±0.2 | 117±1 | 60±0.1 | 900±100 | 20±0.5 | 2±0.1 >12 | H, A, h accessible | 30% ($\mu$) |
| VI Degen | 299±1 | 5TeV >4TeV ($\beta=\pi/2$) <7.5TeV ($\beta=\pi/4$) | 5TeV <∞ ($\beta=\pi/2$) >2.2TeV ($\beta=\pi/4$) | 300±1 | 20 No limit | Pure Higgsino | $\Delta m$ to ±2% | 1.2% |

**Table IV:** *This table summarizes the accuracies expected on SUSY parameters and on the DM estimates. When dominant, the parameter giving the DM error is indicated in the last column.*

In scenarios with heavy sfermions, the neutralino channels have very small cross sections and therefore are challenging to separate from the huge backgrounds. The importance of a very pure b-tag to select the $Z^*$ final states was emphasized.

Mass resolution for the hadronic decays is relevant in two ways. First to separate the $Z^*$ states from the $ZZ$ background. Second, and more important, to measure from the chargino channel the mass difference between the chargino and the LSP.

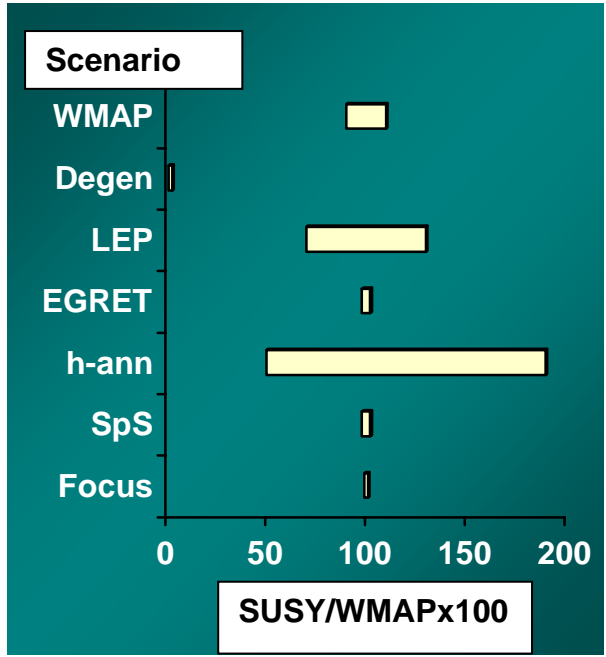

**Figure 11:** *Summary of precisions achieved in this study compared to WMAP.*

Does the chargino analysis require a polarized positron beam? The real issue there is whether one can measure polarisation at the adequate level, meaning at the per mill level given the high statistics involved. We think, as was suggested in [26], that this can be done using $e^+e^- \rightarrow W^+W^-$ since this reaction, completely dominated by the left-handed contribution, provides an adequate analyser to extract the polarisation parameter. One is of course aware that polarized positrons help for other reasons: increase of the overall polarisation, increase in the effective luminosity.

Can a $\gamma\gamma$ collider help? In the EGRET and LEP scenarios it could be the case. For EGRET one expects heavy MSSM Higgses reachable at ILC but this would require operating ILC near 1 TeV while a $\gamma\gamma$ collider operating slightly above 500 GeV could presumably do the job. In the LEP scenario one can improve significantly the accuracy on DM predictions by using the $\gamma\gamma$ collider mode to determine $\tan\beta$.

The capability of ILC to measure the chargino channel in case of mass degeneracy with the neutralino was illustrated with one specific scenario. It seems that ISR final states can be easily separated from $\gamma\gamma$ background events due to a natural suppression of radiation for these events. Our analysis is generic and can be applied to various scenarios recently proposed [32,33]

In summary, although preliminary and incomplete, this analysis has shown that a set of SUSY DM scenarios with heavy sleptons are very challenging and require full use of ILC capabilities.

**Acknowledgements:** We are greatly indebted to A. Djouadi for stimulating discussions and careful reading of this document. Our colleague K. Moenig has provided very useful and critical advice to this work.

# APPENDICES

## I Use of ISR for degenerate cases

In this Appendix:

- we give a simple derivation of ISR formulae valid for annihilations into gauginos
- we derive a simple formula for the $\gamma\gamma$ background, taking into account interferences with FSR.

**ISR for the signal**

Given an annihilation cross section $\sigma(s)$, the differential distribution of ISR is given by:

$$d\sigma/dx = \sigma[(1-x)s]P(x)$$

where $x=k/E$, $P(x)=0.5b[1+(1-x)^2]/x^{1-b}$ and $b$ is an equivalent radiator resulting from an angular integration.

The angular distribution of ISR is given by:
$$dP/d\cos\theta = 2\alpha/\pi \sin^2\theta/(1-\beta^2\cos^2\theta)^2$$

where $\theta$ is the angle of the photon and $\beta$ is the velocity of the beam particles. Integrating over angles one gets:
$$b = 2\alpha/\pi\left[\log(s/m_e^2) - 1\right]$$

For a given value of $k$, one needs to have an angle such that $k\sin\theta > E\theta_{veto} = p_t$ to avoid contamination from $\gamma\gamma$ interactions.

After integration over angles, $b = 2\alpha/\pi \log\left[(1+u)/(1-u)\right]$ taken from 0 to $\cos\theta_{min} = u$ with $\sin\theta_{min} = p_t/k$ and then over $k$, one gets $\sigma_{eff} = b_{eff}\sigma$ with:

$$b_{eff} \simeq 2\alpha/\pi\left\{\log^2 x + \log x\left[\log(s/p_t^2) - 1\right]\right\}$$

taken between $x_{max}=1-4m^2/s$ and $x_{min}=p_t/E$.

Choosing ECM=800 GeV, $m$ =300 GeV, $x_{min}$=5/400, $p_t$=4GeV, then $b_{eff}$ =6.9% to be compared to 46% without angular restriction. The point-like cross section is 150 fb at 800 GeV hence $\sigma_{eff}$~10fb, quite comfortable to measure the radiative cross section. For the wino case this cross section is about twice larger. The generator Susygen confirms this estimate.

**ISR for the $\gamma\gamma$ background**

For this process, it is critical to take into account interference with scattered electrons. Annihilation can be understood as the coherent sum of electron and positron radiation:

$$\left[1/(1-\beta\cos\theta)+1/(1+\beta\cos\theta)\right]^2 \sin^2\theta = 4\sin^2\theta/\left(1-\beta^2\cos^2\theta\right)^2$$

where $\sin^2\theta$ is an helicity factor and $b$ is simply obtained by multiplying this expression by $\alpha/2\pi$.

For $\gamma\gamma$, let's assume that electron (positron) has a scattering angle $\varepsilon$ ($\varepsilon'$) and lets forget about azimuths. One has:

$$\left[1/(1-\cos\theta)-1/(1-\cos\theta+\varepsilon\sin\theta)+1/(1+\cos\theta)-1/(1+\cos\theta-\varepsilon'\sin\theta)\right]^2 \sin^2\theta$$
$$\simeq \sin^2\theta\left[\varepsilon\sin\theta/(1-\cos\theta)-\varepsilon'\sin\theta/(1+\cos\theta)\right]^2$$

Since the two scattered angles $\varepsilon$ and $\varepsilon'$ are uncorrelated, the averaged value is:

$$2\varepsilon^2 \sin^4\theta\left(1+\cos^2\theta\right)/\left(1-\cos^2\theta\right)^2 = 2\varepsilon^2\left(1+\cos^2\theta\right)$$

$b$ is again obtained by multiplying by $\alpha/2\pi$.

What is $\varepsilon$? The transfer to the photon, $q^2$, goes from $q^2_{min}=m_e^2 W^2/s$ to $q^2_{max}=(E\theta_e)^2$ with a differential probability $\sim dq^2/q^2$. Hence $\langle q^2\rangle = q^2_{max}/\log(q^2_{max}/q^2_{min}) = 0.12$ with $W$=10 GeV, where $W$ is the mass of the $\gamma\gamma$ system. This gives $\theta_e$=0.001. With $\varepsilon\sim 0.001$ the suppression is $2\times10^6$ at $\theta=\pi/2$ with respect to the annihilation term. The code BDKRC[25] gives 10 events with 500fb$^{-1}$ and $4\times10^5$ fb for tau pair production corresponding to a suppression of $2\times10^6$ with respect to the annihilation mechanism, in good agreement with above evaluation.

## II Errors for $\chi\chi'$

Neutralino masses can be measured using a threshold scan. A simple method is presented to optimise this 'scan' assuming that two energy points are chosen with an a priori knowledge of the neutralino masses (this method applies equally well to the chargino channels). This knowledge, for example, could come from the lightest chargino analysis.

One assumes that near threshold $\sigma\sim A\beta$, where $\beta^2$ varies like $1-(m+m')^2/s$. One has to determine two parameters: the neutralino mass $m'$ and the cross section constant $A$. The SM background $b$ is known. Assuming that data are taken with $L_2$=500fb$^{-1}$ well above threshold, at an energy $E_2$ defined in table II and that we can spend $L_1$=100fb$^{-1}$ at a lower energy $E_1$, what is the optimal choice? The lowest energy has to be chosen with two criteria in mind:

- provide a significant measurement, i.e. 5 s.d. above background
- maximize the sensitivity to the mass.

The number of events is given by ($A$ includes the detection efficiency): $n_1=L_1(A\beta_1+b)$ and $n_2=L_2(A\beta_2+b)\sim L_2 A\beta_2$, neglecting the background for the highest energy. One requires a significant signal ($k$ s.d.) at the lowest energy: $(n_1-L_1 b)/\sqrt{L_1 b} = k = L_1 A\beta_1/\sqrt{L_1 b}$ hence

$$L_1 A\beta_1 = k\sqrt{L_1 b} \qquad (1)$$

This expression allows to compute $\beta_1$. One can eliminate $A$ and write $n_1-L_1 b=n_2 L_1\beta_1/L_2\beta_2$ which gives:

$$\sqrt{1+(A\beta_1/b)} = k\delta\beta_1/\beta_1 = k\delta(m+m')\left(1/\beta_1^2\right)/(m+m')$$

keeping in mind that $\beta_1 \ll \beta_2$ and that the statistical error on $n_2$ is negigible, this gives:
$$\delta(m+m')/(m+m') = \beta_1^2/k\sqrt{1+(A\beta_1/b)} \qquad (2)$$
As an example, let's take the Focus solution for $\chi_1\chi_2$. One has $A\beta_2$=1.7 fb giving $A$=3.4 fb with $\beta_2$~0.5. From equation (1), with $k$=5, one finds that $\beta_1$=0.03. The corresponding error on $m'$ is 0.3 GeV. Knowing this mass one can in turn determine $A$, i.e. the cross section. This can be done using the high energy point to an accuracy of 3.5% (at this level of precision the uncertainty on the mass is negligible).

In the text, table III shows the accuracies which can be achieved.